\documentclass[lettersize,journal]{IEEEtran}
\usepackage{amsmath,amsfonts}
\usepackage{algorithm2e}
\usepackage{array}
\usepackage[caption=false,font=normalsize,labelfont=sf,textfont=sf]{subfig}
\usepackage{textcomp}
\usepackage{stfloats}
\usepackage{url}
\usepackage{verbatim}
\usepackage{graphicx}
\usepackage{cite}
\usepackage{graphicx} 
\usepackage{amsmath}
\usepackage{amsmath}
\usepackage{amsmath, amssymb}
\usepackage{xcolor}	
\usepackage{amsmath} 

\usepackage{algpseudocode}
\usepackage{graphics}

\usepackage{amsmath, amsthm}

\usepackage[caption=false,font=normalsize,labelfont=sf,textfont=sf]{subfig}

\newcounter{assumption} 

\newcommand{\assumption}[3]{%
	\refstepcounter{assumption}%
	\noindent\hspace{0.5em}\textit{Assumption \theassumption \label{#1}}: #3\par%
}

\newcounter{theorm} 

\newcommand{\theorm}[3]{%
	\refstepcounter{theorm}%
	\noindent\hspace{0.5em}\textit{Theorem \thetheorm \label{#1}}: #3\par%
}

\begin{document}
	
	\title{Enhanced Over-the-Air Federated Learning Using AI-based Fluid Antenna System}
	
	\author{IEEE Publication Technology,~\IEEEmembership{Staff,~IEEE,}
	}
	
	

\author{
    \IEEEauthorblockN{Mohsen Ahmadzadeh\IEEEauthorrefmark{1}, Saeid Pakravan\IEEEauthorrefmark{2}, Ghosheh Abed Hodtani\IEEEauthorrefmark{1}, Ming Zeng\IEEEauthorrefmark{2}, Jean-Yves Chouinard\IEEEauthorrefmark{2}, and Leslie A. Rusch\IEEEauthorrefmark{2}}
    
    \IEEEauthorblockA{\IEEEauthorrefmark{1}Department of Electrical and Computer Engineering, Ferdowsi University of Mashhad, Mashhad, Iran}
    \IEEEauthorblockA{\IEEEauthorrefmark{2}Department of Electrical and Computer Engineering, Laval University, Quebec, Canada}
    
    \IEEEauthorblockA{Email: m.ahmadzadehbolghan@mail.um.ac.ir; saeid.pakravan.1@ulaval.ca; hodtani@um.ac.ir; ming.zeng@gel.ulaval.ca; Jean-Yves.Chouinard@gel.ulaval.ca; leslie.rusch@gel.ulaval.ca}
}

	\maketitle

	\begin{abstract}
		
		This paper investigates an over-the-air federated learning (OTA-FL) system that employs fluid antennas (FAs) at an access point. The system enhances learning performance by leveraging the additional degrees of freedom provided by antenna mobility. We analyze the convergence of the OTA-FL system and derive the optimality gap to illustrate the influence of FAs on learning performance. With these results, we formulate a nonconvex optimization problem to minimize the optimality gap by jointly optimizing the positions of the FAs, the beamforming vector, and the transmit power allocation at each user. To address the dynamic environment, we cast this optimization problem as a Markov decision process and propose the recurrent deterministic policy gradient (RDPG) algorithm. Finally, extensive simulations show that the FA-assisted OTA-FL system outperforms systems with fixed-position antennas and that the RDPG algorithm surpasses the existing methods.		
	\end{abstract}
	
	
	\section{Introduction}\thispagestyle{empty}
	Federated learning (FL) has gained significant traction in communication systems due to its decentralized framework and robust privacy protection measures [1], [2]. Using the computational capabilities of edge devices, FL enables the collective training of a unified global model while ensuring the confidentiality of locally stored sensitive data. This approach is particularly beneficial for various mobile internet of things (IoT) applications, including the internet of drones [3], [4], mobile crowd sensing [5], and other related scenarios. However, implementing FL comes with notable challenges related to communication latency and costs. These challenges can hinder the efficiency and scalability of FL in practical scenarios. To address these issues, over-the-air computation (AirComp) for model aggregation has emerged as an effective solution. AirComp exploits the superposition property of wireless multiple access channels to allow simultaneous data transmission from multiple devices, significantly reducing the overhead involved in traditional aggregation methods [6]. However, over-the-air FL (OTA-FL) model aggregation faces challenges from adverse wireless conditions, particularly in massive mobile IoT scenarios.
	
	To address the challenges of adverse wireless propagation conditions in OTA-FL systems, previous research has extensively explored the integration of reconfigurable intelligent surfaces (RIS) to improve model aggregation reliability [7], [8]. RIS achieves this by reconfiguring wireless channels through passive reflecting elements that adjust their coefficients, effectively steering signals to improve transmission [6]. Although RISs can reshape channel conditions, they are limited by their static positioning and dependency on the surrounding environment, which can hinder performance improvements in dynamic scenarios. To further enhance OTA-FL performance, other studies have investigated advanced beamforming techniques at the receiver, leveraging spatial degrees of freedom to improve signal reception [9]. However, these techniques are also constrained by the fixed positions of receiver antennas, limiting the flexibility of beamforming solutions in dynamic environments.

In contrast, we propose the use of fluid antennas (FAs) in OTA-FL systems to overcome these limitations. Unlike fixed-position antennas (FPAs), FAs possess the unique capability to dynamically manipulate wireless channel conditions through adaptive movement, introducing additional degrees of freedom that can further enhance OTA-FL performance [10]. This adaptability enables FAs to respond in real-time to changing environmental conditions, which is particularly beneficial in mobile IoT contexts where channel characteristics can vary significantly [11], [12]. Previous studies have highlighted the superior performance of FAs over traditional FPAs across various communication systems, including AirComp systems [13], [14], multi-user uplink communications [10], [15], mobile edge computing [16], and covert communication [17]. FAs have also been shown to maximize network sum-rate in multiple-access communication systems through deep reinforcement learning (DRL) [18]. Despite these advances, the integration of FAs into OTA-FL systems remains unexplored.

We propose the integration of FA systems with OTA-FL to enhance convergence performance. We minimize the optimality gap through joint optimization of the beamforming vector, the antenna position vector at the access point (AP), and the transmit power allocation at each user under practical dynamic conditions. We first derive the optimality gap between the actual loss and the optimal loss for OTA-FL to quantify the impact of the beamforming vector and antenna positioning. Based on this convergence analysis, we formulate a non-convex optimization problem aimed at improving learning efficiency. Finally, we reformulate it as a Markov decision process (MDP) and apply DRL techniques for dynamic environments.

	To address the dynamic nature of wireless channels in OTA-FL systems, we introduce the novel integration of FAs with a customized recurrent deterministic policy gradient (RDPG) algorithm. The RDPG algorithm is uniquely designed with actor and critic networks that capture the temporal correlation of state features, enabling real-time decision-making under rapidly changing wireless conditions. This approach not only leverages the flexibility of FAs to dynamically reshape channel environments, but also enhances the adaptability of the learning process by optimizing the antenna positions and beamforming vectors in a dynamic setting. To demonstrate the efficacy of integrated FAs within OTA-FL systems, we conduct extensive simulations comparing the performance of our proposed RDPG algorithm against standard DRL techniques, including soft actor-critic (SAC) and deep deterministic policy gradient (DDPG). Simulation results show RDPG outperforms in performance and stability, highlighting OTA-FL with FAs' superiority over FPAs.
	
	\textit{Notations:} Italicized letters represent scalars, while boldface letters denote vectors. \((\cdot)^T\) is the transpose, \((\cdot)^H\) the conjugate transpose, and \(\mathbb{E}[\cdot]\) the expectation operation. \(|\cdot|\) signifies the magnitude of a scalar or the cardinality of a set. The Euclidean norm of a vector is represented by $\left\| . \right\|$.
	
\section{System Model }
	
 We consider an OTA-FL system comprising \( K \) single-antenna user equipment (UE) devices, denoted as \(\text{UE}_k\), \( \forall k \in \mathcal{K} \triangleq \{1, 2, \ldots, K\} \). The UEs are randomly and uniformly distributed and move dynamically within a designated area of interest, where they collect local data samples. These samples are collaboratively utilized to train a global model at an AP equipped with \( N \) FAs.
	
	\subsection{ OTA-FL Model}
	
	We consider an OTA-FL framework with full participation that executes sequential actions at each iteration \(t\) over \(T\) training rounds as follows:	
	\begin{itemize}
		\item \textbf{Global model broadcast:} The AP broadcasts the current global model \(\boldsymbol{w}_t \in \mathbb{R}^d\) to all UEs, where \(d\) is the dimensionality of the model parameter space.
		\item \textbf{Local model update:} Each UE updates its local model using the gradient descent algorithm as \(\boldsymbol{w}_{k,t} = \boldsymbol{w}_{t} - \gamma \nabla {F}(\boldsymbol{w}_{t}, \mathbf{\mathcal{D}}_k)\), where \(\gamma\) is the learning rate, \(\nabla {F}(\boldsymbol{w}_{t}, \mathbf{\mathcal{D}}_k)\) represents the gradient of the local loss function, and \(\mathbf{\mathcal{D}}_k\) is the local dataset for \(\text{UE}_k\) with a local dataset size denoted by \(|\mathbf{\mathcal{D}}_k| = D\).
		
		\item \textbf{Model aggregation:} 
		Each UE transmits its local model to the AP, which then performs aggregation by averaging to update the global model as:
		\begin{equation}
			\begin{aligned}
				\boldsymbol{w}_{t+1} &= \frac{1}{K} \sum_{k \in \mathcal{K}} \boldsymbol{w}_{k,t}. 
			\end{aligned}
			\label{GD}
		\end{equation}
	\end{itemize}
	The procedure continues iteratively until reaching the maximum specified number of outer iterations.
	
	\subsection{ Communication Model}
	We consider the uploading phase within the OTA-FL system, where each UE synchronously transmits its updated model parameters to the AP. The AP is equipped with an array of FAs, facilitating the adjustment of each FA along a one-dimensional line segment of length $X$. Each FA position is constrained within the interval $[0, X]$ with a minimum distance $X_0$ between adjacent FAs to prevent antenna coupling. The collective locations of all $N$ FAs are represented as the vector $\mathbf{x} = [x_1, \ldots, x_N]^T$, with their movement along one dimension restricted by $x_1 < x_2 < \ldots < x_N$. Time indices are omitted for brevity and clarity in this subsection.
	

The channel between $\text{UE}_k$ and the AP, denoted as $\mathbf{h}_k[\mathbf{x}] \in \mathbb{C}^{N \times 1}$, follows a Rician fading model as:
\begin{equation}
\mathbf{h}_k[\mathbf{x}] = \sqrt{\frac{A_L d_k^{-\alpha_L} \kappa_r}{\kappa_r + 1}} \mathbf{h}_k^{\text{LOS}}[\mathbf{x}] + \sqrt{\frac{A_N d_k^{-\alpha_N}}{\kappa_r + 1}} \mathbf{h}_k^{\text{NLOS}},
\end{equation}
where $\kappa_r$ represents the Rician factor, \(d_k\)
  is the distance between the FAs and $\text{UE}_k$, and $A_L$ and $A_N$ are the path loss at the reference distance for the line-of-sight (LoS) and non-line-of-sight (NLoS) components, respectively. The parameters $\alpha_L$ and $\alpha_N$ denote the path loss exponents for the LoS and NLoS components, respectively. The term $\mathbf{h}_k^{\text{LOS}}[\mathbf{x}]$ represents the LoS component, while $\mathbf{h}_k^{\text{NLOS}}$ denotes the NLoS component. All $\mathbf{h}_k^{\text{NLOS}} \in \mathbb{C}^{N \times 1}$ follow an i.i.d. complex Gaussian distribution with zero mean and unit variance. The LoS component $\mathbf{h}_k^{\text{LOS}}[\mathbf{x}]$ is [13]:
	\begin{equation}
		\mathbf{h}^{\text{LOS}}_{k}[\mathbf{x}] =  [e^{j \frac{2\pi}{\lambda} x_1 \cos(\phi_{k})}, \ldots, e^{j \frac{2\pi}{\lambda} x_N \cos(\phi_{k})} ]^T,
	\end{equation}
	where \(\lambda\) and \(\phi_k\) are the wavelength and the angle of arrival (AoA) of the LoS path, respectively, determined by the location of the UEs in each training round. 
 In this system, we assume each UE moves within an designated area and then transmits model parameters from a stationary position [5]. Moreover, given that the signal path length significantly exceeds the FA movement area, we assume the far field condition between the AP and UEs. Consequently, \( \phi_k \) and \( d_k \) are treated as constants during transmission, regardless of FA positional changes [15], [16]. 
	The AP receives the local model parameters from all UEs in the $t$-th training round as:
	\begin{equation}
		\mathbf{y} = \sum_{k \in \mathcal{K}} {p}_{k}\mathbf{{h}}_{k}[\mathbf{x}]  \boldsymbol{w}_{k} + \mathbf{z},
	\end{equation}
	where, \( p_{k} \) denotes the transmission power factor for the \(k\)-th UE, and \(\mathbf{z} \in \mathbb{C}^{N \times d}\) represents an additive white Gaussian noise (AWGN) matrix with elements following a complex normal distribution \(\mathcal{CN}(0, \sigma^2)\). We consider that the transmission power allocated to each \( \text{UE}_k \) does not exceed the maximum transmission power limit \( p_{\text{max}} \), as [9], [19]: 
	\begin{equation}
		\label{power}
		\frac{1}{d} p_{k}^2 \mathbb{E} \left[ \| \boldsymbol{w}_{k} \|^2 \right] \leq p_{\max}, \quad \forall k \in \mathcal{K}.
	\end{equation}
	The aggregated model parameter vector, \(\boldsymbol{\hat{w}}\), in the \(t\)-th training round is estimated by conducting post-processing on the received signal at the AP as follows:
	\begin{equation}
		\begin{aligned}
			\label{rt}
			\boldsymbol{\hat{w}}& = \frac{\boldsymbol{m}^{H} \mathbf{y}}{K\sqrt{\eta}}  = \frac{1}{K}(\sum_{k \in \mathcal{K}} \frac{1}{\sqrt{\eta}} \boldsymbol{m}^{H} {p}_{k}\mathbf{h}_{k}[\mathbf{x}]  \boldsymbol{w}_{k} + \frac{\boldsymbol{m}^H \mathbf{z}}{ \sqrt{\eta}}),
		\end{aligned}
	\end{equation}
	where, $\boldsymbol{m} \in \mathbb{C}^{N \times 1}$ is the beamforming vector at the AP, and $\eta$ is the scaling factor for signal amplitude alignment.

	\section{Convergence Analysis}	
	To facilitate our convergence analysis, we adopt the following assumptions as discussed in [3], [6], [19]:
	
	\assumption{a1}{assumption1}
	{The global loss function $F(\boldsymbol{w})$ is \( \ell \)-smooth. Namely, for any given model parameters \( \boldsymbol{w}, \boldsymbol{v} \in \mathbb{R}^d \), there exists a nonnegative constant \( \ell \), such that
		\begin{equation}
			F(\boldsymbol{w}) - F(\boldsymbol{v}) \leq (\boldsymbol{w} - \boldsymbol{v})^T\nabla F(\boldsymbol{v}) + \frac{\ell}{2} \|\boldsymbol{w} - \boldsymbol{v}\|^2.
	\end{equation}}
	\assumption{a2}{assumption2} {The loss function satisfies the Polyak-Lojasiewicz inequality, where $F(\boldsymbol{\boldsymbol{w}}^*)$ denotes the optimal global loss value and $\mu > 0$, that is,
		\begin{equation}
			\label{pl}
			\|\nabla F(\boldsymbol{w})\|^2 \geq 2\mu[F(\boldsymbol{w}) - F(\boldsymbol{w}^*)].
		\end{equation}
	}
	\assumption{a3}{assumption3} {The upper limit of the model parameter for \(\text{UE}_k\) is denoted as \( \Gamma \geq 0 \), that is,
		\begin{equation}
			\mathbb{E} \left[ \left\| \boldsymbol{w}_{k} \right\|^2 \right] \leq \Gamma, \quad \forall k \in \mathcal{K}.
		\end{equation}
	}
	\theorm{k}{theorem 1}{Under the conditions outlined in Assumptions \ref{a1}, \ref{a2}, and \ref{a3}, and setting the learning rate to \( 1/\ell \), the optimality gap after \(T\) rounds of training is bounded as follows:
		\begin{equation}
			\begin{aligned}
				\mathbb{E}[ F(\boldsymbol{w}_{T+1})] - F(\boldsymbol{w}^*) &\leq \psi^T\left( \mathbb{E}[F(\boldsymbol{w}_1)] - F(\boldsymbol{w}^*) \right)  \\&+ \sum_{t=1}^{T} \psi^{T-t}\Theta_t=\Phi_T,
			\end{aligned}
		\end{equation}
where, $\Theta_t= \frac{l\Gamma}{2K^2} \sum_{k \in \mathcal{K}} |\frac{1}{\sqrt{\eta}} \boldsymbol{m}^{H} p_{k} \mathbf{h}_{k}[\mathbf{x}] - 1 |^2  + \frac{ld\sigma^2}{2K^2 \eta} \|\boldsymbol{m}^H\|^2$ and $\psi= 1 - \frac{\mu}{l}$.

		\begin{IEEEproof}
			See Appendix. 
	\end{IEEEproof} }

\section{Problem Formulation}  
We enhance learning performance in OTA-FL through the design of FA systems within dynamic environments. According to Theorem \ref{k}, the optimality gap is influenced by the configuration of the beamforming vector, the FA locations, the transmit power factor at each client, and the scaling factor in the training iterations. Thus, we formulate an optimization problem to jointly optimize \(\boldsymbol{m} = [m_1, \ldots, m_N]^T\), \(\boldsymbol{x} = [x_1, \ldots, x_N]^T\), \(\boldsymbol{p} = [p_1, \ldots, p_K]^T\) for all \(k \in \{1, \ldots, K\}\), and the scaling factor \(\eta\), aiming to minimize the total optimality gap as follows:  
\begin{equation}
	\begin{aligned}
		\mathcal{P}_1: &\min_{\boldsymbol{m}, \boldsymbol{x}, \boldsymbol{p}, \eta}\Phi_T \\
		\text{s.t.} &\quad C_1: 0 \leq x_n \leq X, \quad \forall n \in \{1, \ldots, N\}, \\
		& \quad C_2: x_n - x_{n-1} > X_0, \quad \forall n \in \{2, \ldots, N\},\\
		& \quad C_3: \frac{1}{d} p_{k}^2 \mathbb{E} \left[ \| \boldsymbol{w}_{k} \|^2 \right] \leq p_{\max}, \quad \forall k \in \{1,\ldots, K\},\\
		& \quad C_4: \eta > 0,
	\end{aligned}
\end{equation}  
where \(C_1\) constrains the permissible range for FA locations, \(C_2\) enforces a minimum separation distance between adjacent FAs, \(C_3\) sets the maximum power budget for each client, and \(C_4\) ensures the scaling factor is positive.

	The non-convex nature of the objective function and the stochastic nature of the dynamic environment, particularly in massive mobile IoT scenarios, make traditional optimization methods intractable for solving $\mathcal{P}_1$. To tackle this issue, we transform $\mathcal{P}_1$ into an online optimization problem, subsequently reformulating it as an MDP.

	Based on Theorem \ref{k}, the optimality gap at the $t$-th training round, denoted as $\Phi_{t}(\boldsymbol{m}, \boldsymbol{x},\boldsymbol{p}, \eta)$, is bounded as follows:
	\begin{equation}
		\label{tt}
		\begin{aligned}
			\Phi_{t} \leq  \text{$\Phi_{t-1}$} +(\psi ^{t} - \psi ^{t -1})(\mathbb{E}[F(\boldsymbol{w}_1)]  - F(\boldsymbol{w}^*)) +\Theta_{t},
		\end{aligned}
	\end{equation}
	where (\ref{tt}) indicates that when $\psi$ and the initial optimality gap $(\mathbb{E}[F(\boldsymbol{w}_1)] - F(\boldsymbol{w}^*))$ are known, the optimality gap is determined by $\Theta_t$ and the previous optimality gap $\Phi_{t-1}$. Thus, the problem $\mathcal{P}_1$ of minimizing the optimality gap after $T$ communication rounds can be transformed into minimizing $\Theta_t$ in each round. This reformulation is expressed as follows:
	\begin{equation}
		\begin{aligned}
			\mathcal{P}_2: &\min_{\boldsymbol{m}, \boldsymbol{x}, \boldsymbol{p}, \eta}\hspace{0.2cm} \Theta_t \\
			\ \ \ \ &\ \text{s.t. } C_1, C_2, C_3, C_4.
		\end{aligned}
	\end{equation}
	
	Leveraging the zero-forcing structure as discussed in [9] and [6], the minimum \(\Theta_t\) can be determined by considering the following optimal transmit scalar:
	\begin{equation}
		\begin{aligned}
			{p}_{k} = \frac{\sqrt{\eta} \left(\boldsymbol{m}^{H} \boldsymbol{h}_{k}[\boldsymbol{x}]\right)^H}{ \left|\boldsymbol{m}^{H} \boldsymbol{h}_{k}[\boldsymbol{x}]\right|^2 }.
			\label{bm}
		\end{aligned}
	\end{equation}	
	Under the assumption of full participation in FL to adhere to the maximum power budget for each client, the upper bound of $\eta$ must satisfy the following condition:
	\begin{equation}
		\begin{aligned}
			\label{eta}
			& \eta \leq \frac{d p_{\text{max}}\left|\boldsymbol{m}^{H} \boldsymbol{h}_{k}[\boldsymbol{x}]\right|^2}{\mathbb{E} \left[ \| \boldsymbol{w}_{k} \|^2 \right]}, \quad \forall k \in \mathcal{K}.
		\end{aligned}
	\end{equation}
	By applying (\ref{eta}) and (\ref{bm}) in (13), we can rewrite problem $\mathcal{P}_2$ as follows:
	\begin{equation}
		\begin{aligned}
			\mathcal{P}_2: \quad & \min_{\boldsymbol{m}, \boldsymbol{x}} \hspace{0.2cm} \frac{l\sigma^2 \Gamma}{2K^2 p_{\max}} \max_{k \in \mathcal{K}} \frac{\| \boldsymbol{m}_t^H \|^2}{| \boldsymbol{m}_t^H \boldsymbol{h}_k[\boldsymbol{x}] |^2} \\
			& \text{s.t. } C_1, C_2.
		\end{aligned}
	\end{equation}
	
	The aforementioned nonconvex optimization problem presents significant challenges for conventional methods due to dynamic user positions and a time-varying environment, which introduce heterogeneity in each training round. Consequently, we adapt a learning-based algorithm to the different states and identify an appropriate solution.	
	
	\section{Proposed DRL Algorithm}
	To address $\mathcal{P}_2$, we deploy a DRL agent on the AP to learn an optimal decision policy that simultaneously optimizes the beamforming vector $\boldsymbol{m}$ and the FA locations $\boldsymbol{x}$ in each training round in order to minimize $\Theta_t(\boldsymbol{m}, \boldsymbol{x})$. Details of the MDP are:
	\begin{itemize}
		\item \textbf{State Space}:
		The state space at time slot $t$ consists of the distances $d_k$ between the FAs and the $\text{UE}_k$, and the AoA of the LoS paths $\phi_k$, $\forall k \in \mathcal{K}$. The state space can be expressed as: $\boldsymbol{s}_t = [[d_1, \ldots, d_K], [\phi_1, \ldots, \phi_K]]$.
		\item \textbf{Action space}: The action space at each time slot $t$ consists of the beamforming vector and the locations of the FAs. Consequently, the action space at time slot $t$ can be expressed as:
		$
		\boldsymbol{a}_t = [[m_1, \ldots, m_N], [x_1, \ldots, x_N]].
		$	 
		\item \textbf{Reward function}:
		Based on definition on Theorem 1, to minimize $\Theta_t(\boldsymbol{m}, \boldsymbol{x})$, the reward function can be formulated as:
		\begin{equation}
			r(\boldsymbol{s}_t, \boldsymbol{a}_t) = \begin{cases}
				r_1, \ \ \ \ \ \ \ \ \ \ \ \ \ \ \ \ \ \ \ \ \ \ \ \ \ \ \ \ \ \ \|\boldsymbol{m}\| = 0, \\
				r_2 \max_{k \in \mathcal{K}}\left(\frac{\|\boldsymbol{m}\|^2}{|\boldsymbol{m}^H \mathbf{h}_k[\boldsymbol{x}]|^2}\right), \ \ \  \text{otherwise},
			\end{cases}
			\label{reward}
		\end{equation}
		where, the constants $r_1$ and $r_2$ are negative values that require tuning during the simulation process to achieve better convergence. Notably, the reward function is formulated as a negative value. Therefore, by maximizing this reward, the agent effectively minimizes $\Theta_t(\boldsymbol{m}, \boldsymbol{x})$.
	\end{itemize}
 
	Since the action space is continuous, we cannot use model-free value-based DRL algorithms such as deep Q-network (DQN), as they can only handle discrete action spaces. Instead, we utilize policy gradient-based reinforcement learning methods. The DDPG algorithm is a suitable off-policy actor-critic approach capable of managing continuous action spaces. However, the fully-connected deep neural networks (DNNs) employed in conventional DDPG are inadequate for capturing the temporal patterns of environmental dynamics, such as user mobility [20]. Therefore, we adjust the RDPG approach by incorporating long short-term memory (LSTM) into the DDPG architecture to exploit temporal state patterns and adapt continuously to environmental dynamics.
	
	The proposed RDPG algorithm uses four neural networks: an actor network (policy network) denoted by \(\pi_{\phi}\) with parameter \(\phi\), which determines actions \(\boldsymbol{a}_t = \pi_{\phi}(\boldsymbol{s}_t) + \xi\) based on states \(\boldsymbol{s}_t\), where \(\xi\) is a random process added to actions for exploration; a critic network (Q-network) with parameters \(\theta\) that computes Q-values \(Q_{\theta}(\boldsymbol{s}_t, \boldsymbol{a}_t; \theta)\) for state-action pairs; a target actor network, which is an older version of the actor network; and a target critic network, which is an older version of the critic network.
	
	We minimize the optimality gap by maximizing the expected reward \( r(\boldsymbol{s}_t, \boldsymbol{a}_t) \) in each training round. The goal of the RDPG, given the state \( \boldsymbol{s}_t \) and action \( \boldsymbol{a}_t \), is to identify a policy that maximizes the expected cumulative reward, defined as:
	\begin{equation}
		\pi^* = \arg \max_{\pi} \mathbb{E}_{\boldsymbol{s}_t, \boldsymbol{a}_t} \left[ \sum_{t=0}^{\infty} r(\boldsymbol{s}_t, \boldsymbol{a}_t) \right].
	\end{equation}
	To achieve this, the actor network is optimized based on the gradient of the objective function \( J(\phi) \) as follows:
	\begin{equation}
		\nabla_{\phi} J(\phi) = \mathbb{E} \left[ \nabla_{a_t} Q_{\theta_1}(\boldsymbol{s}_t, \boldsymbol{a}_t) \bigg|_{a_t=\pi_{\phi}(\boldsymbol{s}_t)} \nabla_{\phi} \pi_{\phi}(\boldsymbol{s}_t) \right].
	\end{equation}
	The critic network is trained to minimize the loss function relative to the target value \( Y_t \), defined as:
	\begin{equation}
		Y_t = r_t + \gamma Q_{\theta_i'}(\boldsymbol{s}_{t+1}, \pi_{\phi'}(\boldsymbol{s}_{t+1}) + {\xi}).
		\label{eq:target}
	\end{equation}
	The proposed RDPG method is described in Algorithm \ref{algorithm}.
	\RestyleAlgo{ruled}
	\SetNlSty{textbf}{}{:} 
	\begin{algorithm}
		\SetAlgoLined
		\textbf{Initialize}: experience replay memory \( M \), mini-batch size \( H \), the actor network \( \pi_{\phi} \), the critic network \( Q_{\theta} \) with random values, and create the target networks by setting \( \theta' \leftarrow \theta \) and \( \phi' \leftarrow \phi \).
		
		\textbf{Set}: Set \( E \) and \( T \) as the maximum number of episodes and episode length, respectively.\\
		
		\For{each episode $e:E{}$}{%
			Initialize the environment state $s_0$, and the exploration noise $\xi$;\\
			\For{$t = 1:T$}{%
				Receive $s_t$ from the environment;\\
				Obtain $\boldsymbol{a}_t = \pi_{\phi}(\boldsymbol{s}_t) + \xi$ from the actor network and re-shape it; \\ 
				Obtain $r_t$ based on equation \eqref{reward};\\
				Observe the new state, $\boldsymbol{s}_{t+1}$;\\
				Store transition $(\boldsymbol{s}_t, \boldsymbol{a}_t, r_t, \boldsymbol{s}_{t+1})$ into $M$;\\ 
			}
			Randomly sample a $H$ mini-batch of transitions from $M$;\\
			Compute the target function $Y_t$ according to \eqref{eq:target};\\
			Update the actor and critic networks using the Adam optimizer. \\
			Soft update the target actor and target critics with \(\tau \in [0,1]\), as the soft update coefficient:
			\[
			\phi' \leftarrow \tau \phi + (1 - \tau)\phi', \quad
			\theta' \leftarrow \tau \theta + (1 - \tau)\theta'
			\]}	
		\caption{The RDPG Algorithm}
		\label{algorithm}
	\end{algorithm}
 
	\subsection{Computational Complexity Analysis}
 
The computational complexity of a DRL network, such as the proposed RDPG algorithm, consists of both the action selection and training processes [21], [22]. The architecture comprises one actor network and one critic network, each with $\mathcal{U}$ hidden layers containing $\mathcal{L}$ neurons per layer. The action selection complexity, which refers to generating network output for a given input, can be derived from the size of the consecutive layers. For the actor network, this is expressed as: 
$\mathcal{J} \times (|S| + |A|) \times \mathcal{L}$ for the input and first layer, 
$\mathcal{L}^2$ for the successive hidden layers, and 
$\mathcal{L} \times |A|$ for the output layer, where $\mathcal{J}$ represents the previous trajectory length, $|S|$ denotes the state dimension, and $|A|$ represents the action dimension.
 For the critic network, the production of the consequence layers is 
	\(\mathcal{J} \times (|S| + 2 \times |A|) \times \mathcal{L}\) for the input and first layer, 
	\(\mathcal{L}^2\) for the successive hidden layers, and 
	\(\mathcal{L} \times |A|\) for the final connection. Here, \(|S|\) and \(|A|\) denote the dimensions of the agent state and action spaces, respectively. Thus, the action selection complexity for proposed method is \(\mathcal{O}(\mathcal{L}^2)\). 

During the training process, the computational complexity of RDPG is determined by the number of network edges, calculated as 
$I \times C + C^2 + C \times O$, where $I$ is the input size, $C$ is the number of neurons, and $O$ is the output size [22]. The complexity for the actor and critic networks can be further refined as: 
$(H |S| \mathcal{L} + H \mathcal{L}^2 + H \mathcal{L} |A|)$ and 
$(H (|S| + |A|) \mathcal{L} + H \mathcal{L}^2 + H \mathcal{L})$, respectively, where $H$ denotes the batch size. Consequently, the overall training complexity for the RDPG is $\mathcal{O}(H \mathcal{L}^2)$. Comparatively, the computational complexity of other DRL algorithms, such as SAC and DDPG, is expressed as 
$\mathcal{O} \left( \left( \sum_{\mathcal{N}=1}^{\mathcal{N}_l} C_{\mathcal{N}} C_{\mathcal{N}-1} \right) H \mathcal{N}_e \right)$, where $\mathcal{N}_l$ is the number of layers, $C_{\mathcal{N}}$ is the number of neurons per layer, and $\mathcal{N}_e$ is the total number of episodes [21]. 

\section{Simulation Results}

	We provide numerical results illustrating how combining FA arrays with the proposed RDPG algorithm can improve OTA-FL learning performance. We assume the distances between users and the AP are independent and uniformly distributed in the range \([20, 100]\) meters, and the AoAs are uniformly distributed over \([-\pi/2, \pi/2]\) radians. The parameters for the FA arrays are set with \( X_0 = 0.5 \lambda \) and \( X = 8 \lambda \). The Rician factor is \(\kappa_r = 10\), the path loss constants are \(A_L = A_N = -2.14\) dB, the path loss exponents are \(\alpha_L = \alpha_N = 2.09\), and \(\lambda\) is set to 1 for simplification.

 The RDPG algorithm is configured with a learning rate of 0.0005, a replay buffer size of $10^4$, a batch size of 64, a soft update parameter of 0.001, and a discount factor of 0.9. For performance evaluation, we compare the FA algorithm to FPA using a predetermined location vector \(\boldsymbol{x} = \left[ \frac{X}{N+1}, \ldots, \frac{NX}{N+1} \right]^T\), and assess the proposed RDPG algorithm against conventional DRL algorithms SAC [23] and DDPG [22].
	Learning performance is evaluated by computing the average rewards over 100 episodes, which is determined at episode \( e \) by employing the \({R_{\text{avg}}(e)} = \frac{1}{100} \sum_{i=e-100}^{e} R_i\), where \( R_i \) signifies the mean reward of episode \( i \).
	\begin{figure*}[!]
		\centering
		\includegraphics[width=17.25cm, height=4.665cm]{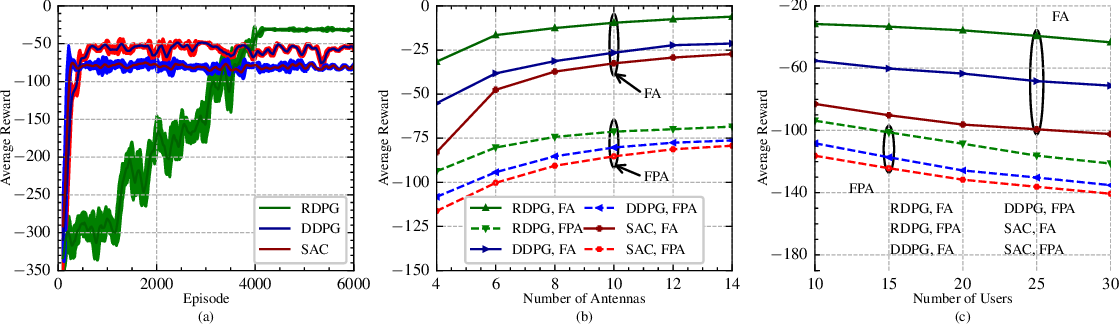}
		\caption{Comparison of DRL Algorithms in FA and FPA systems: (a) training episodes for $K=10$ and $N=6$; (b) antenna numbers; and (c) client numbers.}
		\label{fig:episode}
	\end{figure*}
	

 Fig. \ref{fig:episode} (a) demonstrates the convergence characteristics of different DRL algorithms, depicting the average rewards with solid curves and showing the standard deviations as shaded regions. The RDPG exhibits higher average rewards and lower variance compared to standard DRLs, demonstrating superior performance and improved stability in dynamic environments.
	

 To evaluate the proposed algorithms with different numbers of antenna, we kept the number of clients fixed and varied the antenna count in both FA and FPA scenarios. As depicted in Fig. \ref{fig:episode} (b), the average reward performance of all DRL methods improves with increasing \(N\), although this improvement diminishes as \(N\) continues to increase. Furthermore, due to the increased degrees of freedom provided by antenna adjustments in FA systems, FAs consistently outperform FPAs at all values \(N\). Moreover, RDPG demonstrating superior performance over other DRL algorithms.
	
	Fig. \ref{fig:episode} (c) provides a detailed comparison of the performance of FAs and the RDPG algorithm across varying numbers of users. As the number of users increases, there is a noticeable decrease in performance for both FA and FPA scenarios. This decline is attributed to the increased challenge of optimizing the beamforming vector and the antenna position vector of the AP in the presence of more dynamic users. Despite these challenges, FAs consistently outperform FPAs in all tested scenarios, highlighting the efficacy of FAs in enhancing OTA-FL system performance. Moreover, the RDPG algorithm consistently exhibits superior performance compared to other optimization methods in mitigating the adverse effects of dynamic user dynamics on system performance.

	\section{Conclusion}
	We demonstrated the integration of FAs into AP to improve the performance of OTA-FL systems. Our convergence analysis highlighted the significant impact of FA positions and the beamforming vector on the optimality gap. We addressed this issue with a non-convex optimization problem and proposed the RDPG algorithm for real-time optimization. Through simulations, we demonstrated that the OTA-FL system enhanced by FAs outperformed conventional FPAs systems. Moreover, RDPG demonstrates superior performance and stability compared to existing methods, validating its effectiveness in dynamic environments.
	
	\appendix
	In the $t$-th communication round, based on (\ref{GD}) and (\ref{rt}), the global model update can be expressed as follows: 
	\begin{multline}
			\boldsymbol{\hat{w}}_{t+1} = \frac{1}{K} \sum_{k \in \mathcal{K}} \boldsymbol{w}_{k,t}  + \mathbf{e}_t 
			= \frac{1}{K} \sum_{k \in \mathcal{K}}(\boldsymbol{w}_{t} - \gamma \nabla {F}(\boldsymbol{w}_{t}, \mathbf{\mathcal{D}}_k)) \\+  \mathbf{e}_t		
			=\boldsymbol{w}_{t} - \gamma (\nabla {F}(\boldsymbol{w}_t) - \frac{1}{\gamma}\mathbf{e}_t),
		\label{wt}
	\end{multline}
	where $\nabla F(\boldsymbol{w}_t) = \frac{1}{K} \sum_{k \in \mathcal{K}} \nabla F_k(\boldsymbol{w}_t, \mathcal{D}_k)$ represents the global gradient, and $\mathbf{e}_t = \boldsymbol{\hat{w}}_{t+1} - \boldsymbol{w}_{t+1}$ denotes the model aggregation error caused by wireless communication.  
Taking the expectation of (\ref{wt}) and considering \eqref{GD} and \eqref{rt}, with $\eta = \frac{1}{l}$, we derive:
	\begin{equation}
		\begin{aligned}
			\label{con2}
			\mathbb{E}[{F}(\boldsymbol{w}_{t+1})] &\leq \mathbb{E}[{F}(\boldsymbol{w}_t)] - \frac{1}{2l} \| \nabla {F}(\boldsymbol{w}_t) \|^2 + \frac{l}{2} \mathbb{E}[\|\mathbf{e}_t\|^2]. 
		\end{aligned}
	\end{equation}	
	Based on (\ref{GD}) and (\ref{rt}), $\mathbb{E}[\| \mathbf{e}_t\|^2]$, is bounded as follows:
	\begin{multline}
		\label{e_t2}
		\mathbb{E}[\| \mathbf{e}_t\|^2] = \mathbb{E}[\|\boldsymbol{\hat{w}}_{t+1} - \boldsymbol{w}_{t+1}\|^2] =\\
			 \frac{1}{K^2} \sum_{k \in \mathcal{K}} |\frac{1}{\sqrt{\eta}} \boldsymbol{m}^{H} p_{k} \mathbf{h}_{k}[\mathbf{x}] - 1 |^2 \mathbb{E}[\|\boldsymbol{w}_{k,t}\|^2] +\frac{d\sigma^2}{K^2\eta} \|\boldsymbol{m}^H\|^2 \\
			\stackrel{\text{a}}{\leq}  \frac{\Gamma}{ K^2} \sum_{k \in \mathcal{K}} |\frac{1}{\sqrt{\eta}} \boldsymbol{m}^{H} p_{k} \mathbf{h}_{k}[\mathbf{x}] - 1 |^2    + \frac{d\sigma^2}{K^2 \eta} \|\boldsymbol{m}^H\|^2,
	\end{multline}
	where (a) follows from Assumption \ref{a3}, which defines the upper bound of the local model parameters.  

By employing Assumptions (\ref{a2}) and (\ref{e_t2}) and subtracting \( F(\boldsymbol{w}^*) \) from both sides of (\ref{con2}), we obtain:
	\begin{multline}
		\label{con3}
		\mathbb{E}[F(\boldsymbol{w}_{t+1})] - F(\boldsymbol{w}^*) \leq (1-\frac{\mu}{l})(\mathbb{E}[F(\boldsymbol{w}_t)] - F(\boldsymbol{w}^*))+\\ \frac{l\Gamma}{2K^2} \sum_{k \in \mathcal{K}} |\frac{1}{\sqrt{\eta}} \boldsymbol{m}^{H} p_{k} \mathbf{h}_{k}[\mathbf{x}] - 1 |^2  + \frac{ld\sigma^2}{2K^2 \eta} \|\boldsymbol{m}^H\|^2.
	\end{multline}
	By recursively applying (\ref{con3}) and using the definitions of $\Theta_t$ and $\psi$ in Theorem 1, the cumulative optimality gap is: 
	\begin{multline}
			\mathbb{E}[F(\boldsymbol{w}_{T+1})] - F(\boldsymbol{w}^*) \leq \psi(\mathbb{E}[F(\boldsymbol{w}_{T})] - F(\boldsymbol{w}^*) + \Theta_{T} \\
			\leq \psi(\psi(\mathbb{E}[F(\boldsymbol{w}_{T-1})] - F(\boldsymbol{w}^*)) + \Theta_{T-1}) + \Theta_{T}) \\
			\leq \ldots 
			\leq \psi^T(\mathbb{E}[F(\boldsymbol{w}_1)] - F(\boldsymbol{w}^*)) + \sum_{t=1}^{T} \psi^{T-t} \Theta_t.
	\end{multline}
	This completes the proof of Theorem \ref{k}.

	\section{References Section}


	\vfill
	
\end{document}